# A One-Dimensional Coordination Polymer, BBDTA·InCl$_4$; Possible Spin-Peierls Transition with High Critical Temperature of 108 K

Wataru Fujita

Research Center for Materials Science, Nagoya University, Furo-cho, Chikusa-ku, Nagoya 464-8602, Japan.

Ryusuke Kondo and Seiichi Kagoshima

Department of Basic Science, The University of Tokyo, Komaba 3-8-1, Meguro-ku, Tokyo 153-8902, Japan.

Kunio Awaga

Department of Chemistry, Nagoya University, Furo-cho, Chikusa-ku, Nagoya 464-8602, Japan.

**Abstract**

We have studied the crystal structure and magnetic properties of the organic radical cation salt, BBDTA•InCl$_4$. This material formed a one-dimensional coordination polymer, whose structure was characteristic of inorganic spin-Peierls materials. Magnetic measurements indicated the spin-Peierls transition like behavior at 108 K, which was higher than those typically observed for the other organic spin-Peierls materials. The structural aspects of the lattice distortion from X-ray diffraction measurements at 50 K have been discussed.

The spin-Peiels (SP) transition occurs in antiferromagnetic Heisenberg $S$ - 1/2 chains in the presence of strong magnetoelastic coupling with three-dimensional lattice vibrations.[1] Below the SP transition temperature ($T_{sp}$), a lattice dimerization occurs which increases progressively as the temperature is lowered, together with the concomitant appearance of an energy gap in the spectrum of magnetic excitations. This gap separates a nonmagnetic singlet ground state from a triplet of excited states. The SP transitions has been found in several organic compounds, such as TTF-$M$S$_4$C$_4$(CF$_3$)$_4$ [$M$ = Cu ($T_{sp}$ = 12.4 K), Au ($T_{sp}$ = 2.06 K)],[2] MEM•(TCNQ)$_2$ ($T_{sp}$ = 20 K),[3] $p$-CyDOV ($T_{sp}$ = 15 K),[4] (DMe-DCNQI)$_2$$M$ [$M$ = Ag ($T_{sp}$ = 86 K), Li ($T_{sp}$ = 52 K)].[5] Recently some inorganic compounds showing the SP transition like behaviors, CuGeO$_3$,[6] TiO$X$ ($X$ = Cl, Br),[7,8] NaTiSi$_2$O$_6$,[9] Sb$_2$O$_2$VO$_3$,[10] *etc*, have been reported.

Cyclic thiazyl radicals have attracted much interest in view of their potential use as building blocks in the synthesis of organic conductors and magnets.[11-18] The neutral radical, β-$p$-NC-(C$_6$F$_4$)-(CNSSN), exhibits weak ferromagnetism with a critical temperature of 35.5 K.[14] The cation radical salts, 1,3- and 1,4-[(S$_2$N$_2$C)C$_6$H$_4$(CN$_2$S$_2$)][$X$] ($X$ = I, Br)



and [(S$_2$N$_2$C)-(CN$_2$S$_2$)][*X*] (*X*= I), have shown semi-metallic behavior and metal-insulator transitions.[12] Our target molecule in this paper is the monocationic state (*S* = 1/2) of benzo[1,2-d:4,5-d']bis[1,3,2]dithiazole (BBDTA), whose molecular structure is presented in Scheme 1. Recently we have found a ferromagnetic ordering state below 7.0 K in the salt, γ-BBDTA•GaCl$_4$, in which GaCl$_4^-$ is a non-magnetic counter anion.[18] In the present work, we report the crystal structure and magnetic properties of the cyclic thiazyl radical derivative, BBDTA•InCl$_4$, which has formed a one dimensional coordination polymer structure and exhibits the characteristic behavior of the SP transition at 108 K. In addition, we propose the possible structural models for lattice dimerization in this material below the transition temperature.

BBDTA•InCl$_4$ was prepared by the reaction of BBDTA•Cl and InCl$_3$ in CH$_3$CN under a nitrogen atmosphere.[19,20] Dark greenish crystals of this material were obtained by recrystallization from a hot acetonitrile solution. Crystal data are as follows:[21] C$_6$Cl$_4$H$_2$In$_1$N$_2$S$_4$ (*M* = 528.037), dark greenish block, 0.10 × 0.20 × 0.20 mm$^3$, orthorhombic, space group *Cmcm a* = 9.0400(4), *b* = 10.4900(5), *c* = 13.9240(8) Å, *V* = 1320.41(11) Å$^3$, *Z*



= 3, $\rho_{calcd}$ = 1.831 g cm$^{-1}$, $T$ = 293 K, $2\theta_{max}$ = 27.92°, $R_1(I > 2\sigma)$ = 0.0433, $wR^2$ = 0.1174 (all data), $S$ = 1.191.  It crystallizes in the orthorhombic *Cmcm* space group, in which a quarter of the BBDTA•InCl$_4$ unit is crystallographically asymmetric.  Figure 1(a) depicts the geometry around the indium atom in this material.  The radical cation is coordinated to the indium atom as a ligand.  The indium atom exhibits a distorted octahedral geometry and is surrounded by four chloride atoms and two nitrogen atoms of BBDTA$^+$ radical cations, in contrast to the tetrahedral environment of the gallium atom in BBDTA•GaCl$_4$.[18]  The In—Cl bond lengths (2.3916 (3) and 2.4398 (4) Å) in BBDTA•InCl$_4$ are similar to those in other reported indium chloride complexes.[22]  The two BBDTA$^+$ cations occupy a *cis* positions with interatomic In—N distances of 2.586 (5) Å, which are longer than those reported for other In—N complexes, on average 2.25 Å.[23]  This indicates that the coordination bond in this material is weaker than those of other analogous another complexes.  In addition, there are some short interatomic distances which are substantially less than the sum of the van der Waals radii between the ligand species.  An interatomic S•••Cl short distance of 3.2886 (11) Å was observed between BBDTA and Cl, as shown by



broken lines in Fig. 1a.  The S•••Cl interatomic contact may be close due to the long In—N coordination bond.  Furthermore, there are short interatomic N—N distances of 3.052 (8) Å (red solid lines) between the BBDTA$^+$ ligands.  We can expect a strong antiferromagnetic interaction between the BBDTA$^+$ cations in the chain.  In the molecular configuration there is large overlap between the magnetic orbitals of the neighbor BBDTA+ cations in the chain, of whose the singly-occupied molecular orbital (SOMO) has a larger population on the nitrogen atoms.[18]  Figure 1b shows the crystal structure of BBDTA•InCl$_4$ at room temperature.  A BBDTA$^+$ radical cation bridges the distance between the InCl$_4^-$ anions, resulting in a one dimensional coordination chain along the *c* axis.  In this crystal, a 1D antiferromagnetic network is formed via the nitrogen-nitrogen short contact.  In the interchain arrangement, there is a short contact between the sulfur atoms on the dithiazolyl ring (3.559 (2) Å), as shown by the broken lines in Fig. 1(b).

Figure 2 depicts the temperature dependence of the paramagnetic susceptibility, $\chi_p$, for BBDTA•InCl$_4$ in the temperature range of 2—400 K.  The value of $\chi_p$ was obtained from the dc susceptibilities, subtracting a diamagnetic susceptibility of $-3.44 \times 10^{-4}$ emu



mol$^{-1}$. We adopt BBDTA•InCl$_4$ as the molar unit. The value of $\chi_p T$ at 400 K is 0.336 emu K mol$^{-1}$, which is smaller than the ideal value, 0.375 emu K mol$^{-1}$, for non-interacting $S$ - 1/2 spins of $g$ = 2.[24] This fact suggests dominance of intermolecular antiferromagnetic interactions in the radical cations. $\chi_p$ shows a gradual increase, decreasing with temperature, and makes a broad maximum at ca. 170 K. This magnetic behavior is characteristic of low dimensional magnetic materials. $\chi_p$ decreases abruptly at 110 K, suggesting a magnetic phase transition. Below this temperature this material shows a gap-like magnetic behavior. This behavior shows no thermal hysteresis, suggesting a second order phase transition. Below 50 K $\chi_p$ again increases, decreasing with temperature. This behavior is due to Curie term from lattice defects or magnetic impurities.

We estimated the exchange coupling constants of the intermolecular antiferromagnetic interactions in the high and low temperature phases. In the high temperature region above 110 K, the best fit of the experimental data was obtained, using the Bonner-Fischer equation[25] with the parameters; $C$= 0.375 emu K mol$^{-1}$ (fixed) and $J/k_B$= -135 K, where $C$ is the Curie constant, $J$ is the intrachain exchange coupling constant, and



$k_B$ is the Boltzmann's constant. As shown in Fig. 2, the magnetic data are well reproduced by the theoretical curve. This indicates that the present material is a typical $S - 1/2$ 1D magnet, as expected from the structure. In the low temperature region the rough estimate for the spin-gap value was obtained by fitting $\chi_p$ to the following equation:[26]

$$\chi_p = \alpha / T \cdot \exp(-\Delta / k_B T) + C_{imp} / T,$$

where $\alpha$ is a constant value corresponding to the dispersion of the excitation energy, $\Delta$ is the magnitude of the spin gap, and $C_{imp}$ is related to Curie components of lattice defects or impurities at low temperature. The value of the spin-gap, $-\Delta / k_B$, obtained from the fitting was 460 K. This magnetic behavior is characteristic of materials exhibiting the spin-Peierls (SP) distortion.

In order to estimate lattice deformation in BBDTA•InCl$_4$ below the transition temperature, we carried out X-ray diffraction measurements of a single crystal in the low temperature region.[27] We did not observe superlattice reflections in the crystal at 50 K, suggesting no significant change in the size of the unit cell below the transition temperature. Lattice dimerization does not always give superlattice reflections because two BBDTA$^+$ molecules are included in the c axis of the unit cell in this material.  We found at 50K several Bragg reflections, (0 5 -7), (3 4 6), (3 -4 0), (-3 4 -6), (3 0 5), and (-3 0 -9) which were not found in the high temperature phase above 110K because of the extinction rules for a *C*-centered unit cell and a *c*-glide plane perpendicular to *b*.[28]  The appearance of these reflection peaks suggests symmetry changes of the crystal below the transition temperature.

    We can propose at least two candidates for the dimerization aspect with no significant change of the unit cell and an absence of the two symmetry operations mentioned above.  Since the In—N coordination bond was expected to be weak judging from the room temperature structure, we assume that changes with the N—In—N bond angle, the In—N bond length, or both structural parameters have occurred at low temperature.  Figure 3 depicts two dimerization possibilities of the coordination chain structure in BBDTA•InCl$_4$ at 50 K.  As shown in Fig. 3a, when only the upper N—In—N bond angles decrease, the corresponding intermolecular N•••N distances may shorten.  In this situation, the BBDTA$^+$ cations can form a dimeric structure.  The other possible

structure is shown in Fig. 3b.  If the BBDTA cations approach the indium atoms in the uppermost part of the chain, that is, if the upper In—N coordination bonds shorten and the lower bonds elongate, it may result in dimerization of the BBDTA cations.  In both of these structural predictions, the structural changes occur in the *b-c* plane.  Hence, it can be expected that some kind of an anomaly is found in the lattice parameters of the *b* and *c* axes around the transition temperature.

We have measured the temperature dependence of the lattice parameters.  The lattice constants of the *a*, *b*, and *c* axes are shown in Fig. 4 as a function of temperature. They are normalized, using their value at 295 K.  Since discontinuous change of these values is not observed, this phase transition may have a second-order character.  The value of all the axes gradually decrease, decreasing with the temperature down to 108 K.  The value of the *b* and *c* axes showed a slight increase and a further decrease below said temperature, respectively.  This temperature dependence is consistent with the above structural predictions.  It seems that the characteristic behavior of the SP transition in the material correlates with nature of the In—N coordination bond.  In order to obtain detailed information on the structural change, we need to carry out X-ray structural analysis at around 50 K.

In summary we have reported the SP transition of the organic radical based magnet, BBDTA•InCl$_4$ at 108 K, which is higher than those typically observed for the other organic SP materials. This material formed a one-dimensional coordination polymer composed of alternating BBDTA$^+$ and the indium chloride anion units. This structural character is common to those of inorganic SP materials rather than traditional organic SP materials composed of stacking of planar molecules. We believe that this high transition temperature of 108 K is due to the long In—N bond length observed in the structure of our coordination polymer, which bring a certain amount of structural flexibility and therefore may diminish the loss of lattice energy in the phase transition. Below the transition temperature we observed some Bragg reflections that did not appear at the high temperature phase in the X-ray diffraction measurements, probably suggesting that lattice dimerization took place. Based on the symmetry changes of the crystal, we predicted roughly the structural aspect of the dimerized phase. More detailed structural studies below the transition temperature on this material are now being carried out.

**Acknowledgement**

We thank J. M. Goicoechea for checking this manuscript. This work was partially supported by the Naito Foundation and 21st Century COE Program: *Elucidation and Creation of Molecular Functions*.

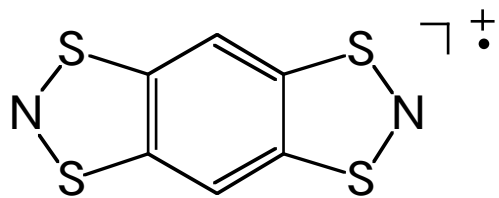

BBDTA⁺

Scheme 1

**Figure Captions**

Figure 1 Crystal structure of BBDTA•InCl$_4$ at room temperature, (a) a coordination environment of an indium atom, (b) crystal packing in BBDTA•InCl$_4$.

Figure 2 Temperature dependence of the paramagnetic susceptibilities, $\chi_p$, in BBDTA•InCl$_4$ under 1 T.

Figure 3 Two structural candidates for the dimeric phase in BBDTA•InCl$_4$ at 50 K. Based on changes of (a) the N—In—N bond angles, or (b) the In—N bond lengths.

Figure 4 Temperature dependence of the lattice parameters in BBDTA•InCl$_4$.

(a)

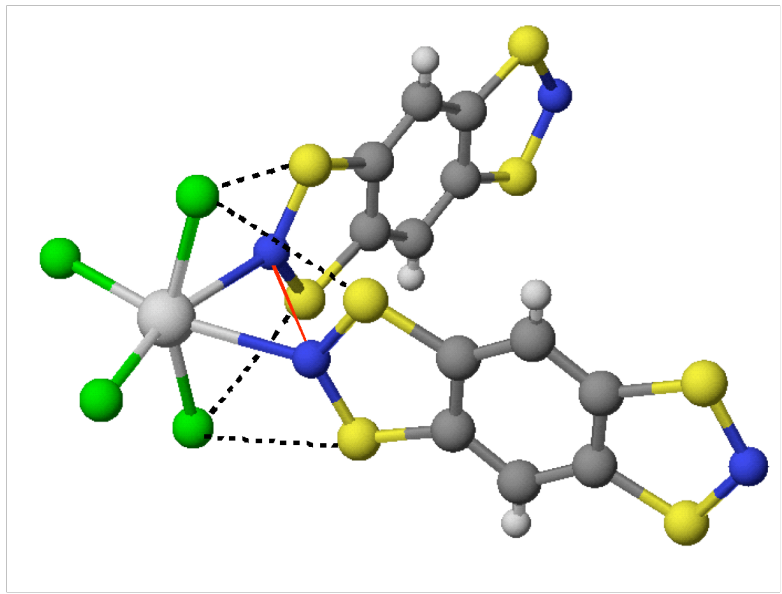

(b)

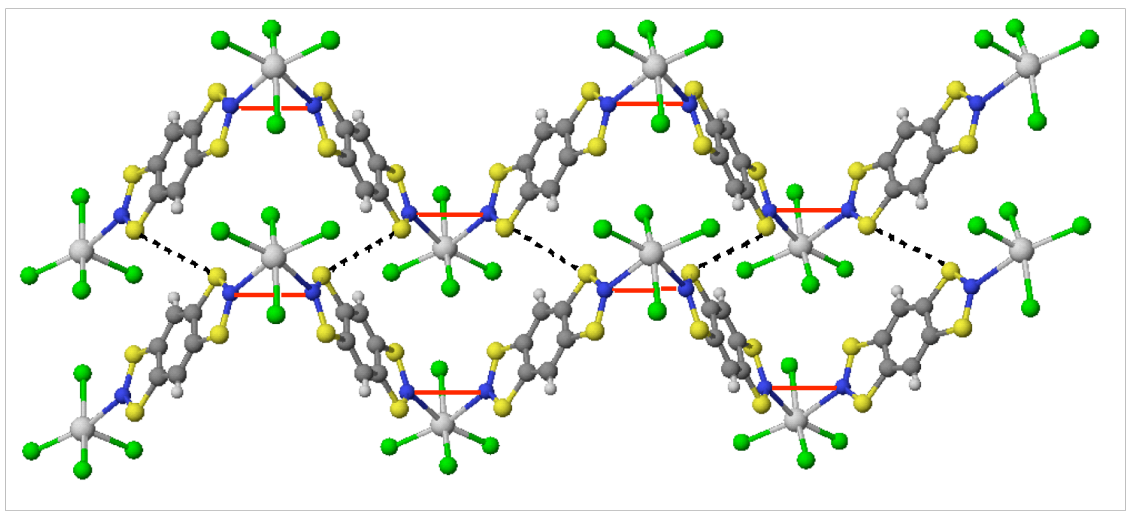

Figure 1 Fujita et al.

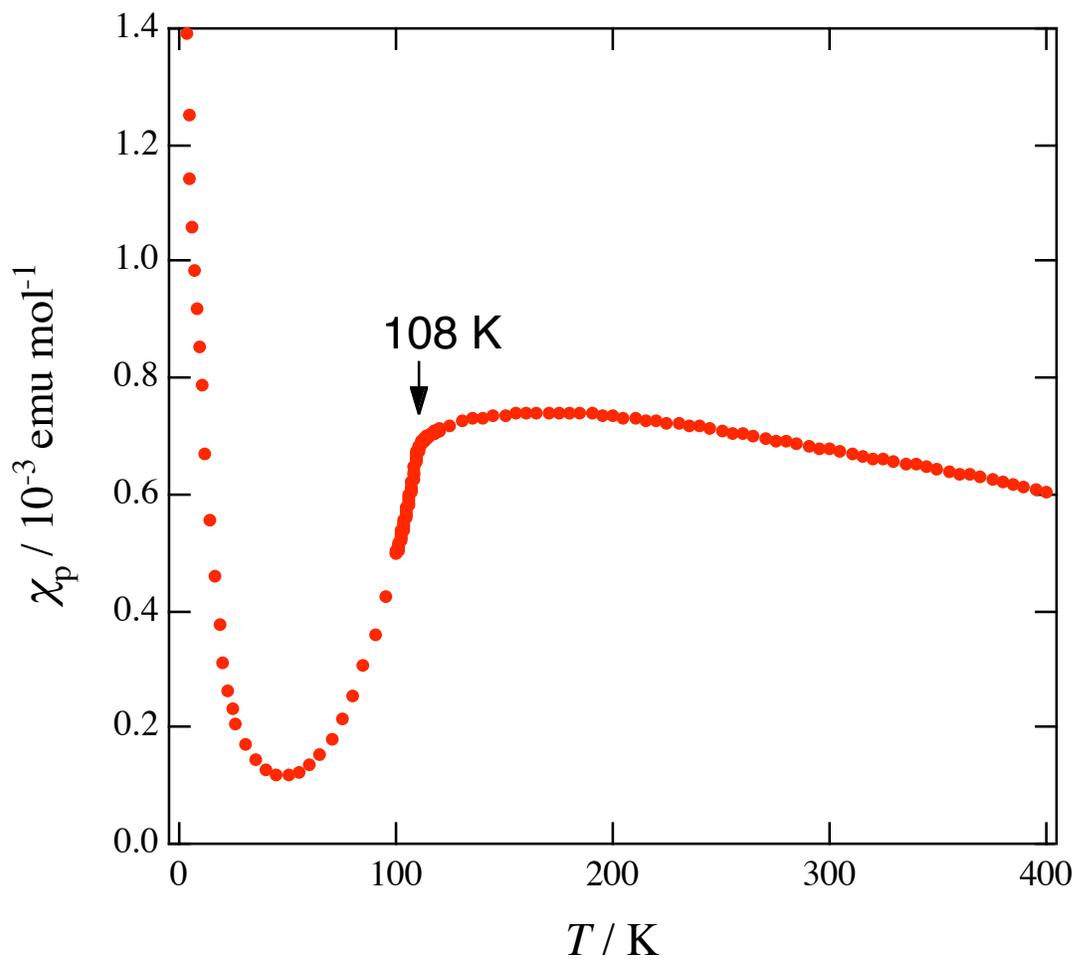

Figure 2 Fujita et al.

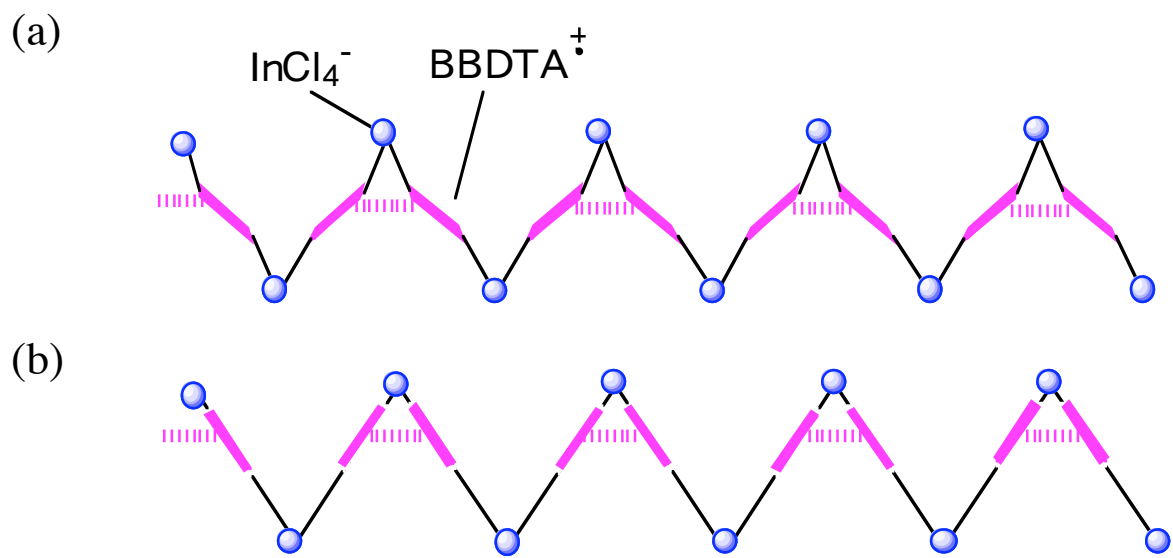

Figure 3 Fujita et al.

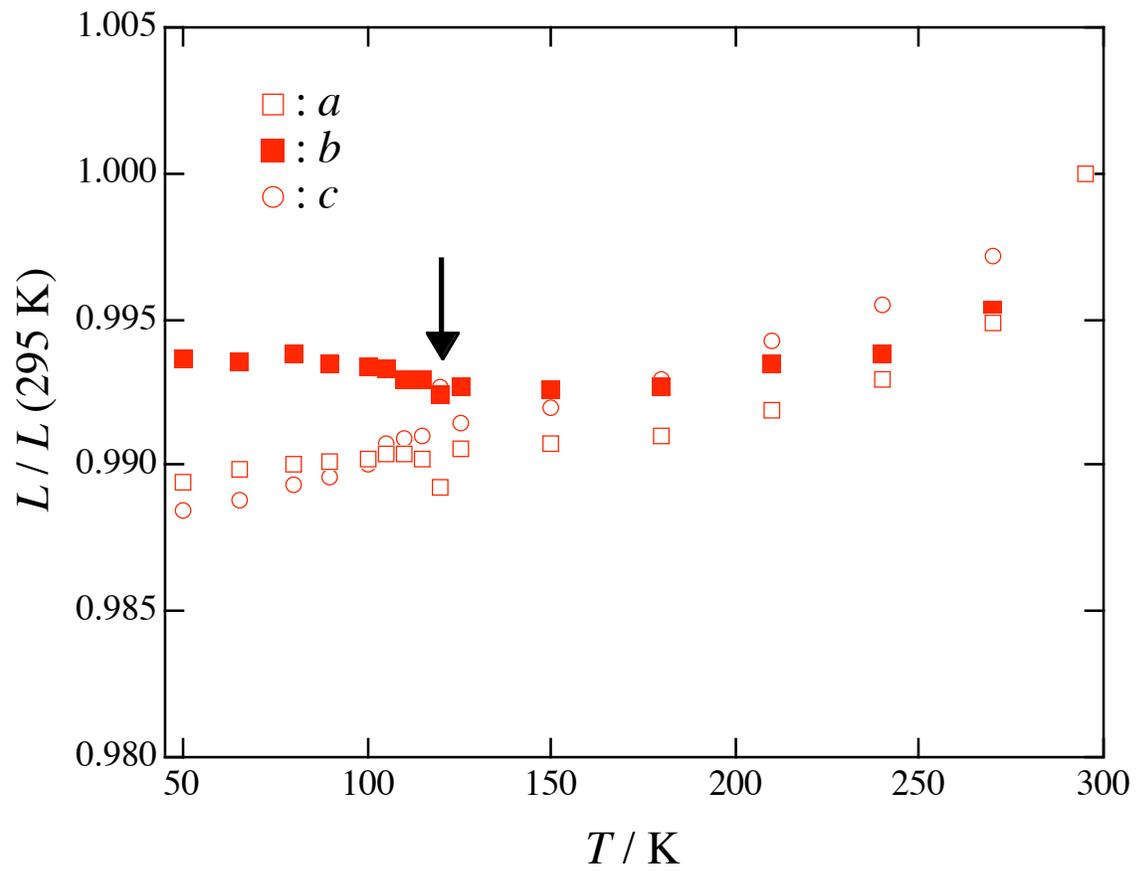

Figure 4 Fujita et al.